\documentclass{article}
\usepackage{frascatiphys}
\usepackage{graphicx}

\def\be{\begin{equation}}
\def\ee{\end{equation}}
\def\bea{\begin{eqnarray}}
\def\eea{\end{eqnarray}}

\def\bea{\begin{eqnarray}}
\def\eea{\end{eqnarray}}
\def\beq{\begin{equation}}
\def\eeq{\end{equation}}
\def\gappeq{\mathrel{\rlap {\raise.5ex\hbox{$>$}} {\lower.5ex\hbox{$\sim$}}}}
\def\lappeq{\mathrel{\rlap{\raise.5ex\hbox{$<$}} {\lower.5ex\hbox{$\sim$}}}}

\begin{document}
\title{ 
THE HIGGS AND THE EXCESSIVE SUCCESS OF THE STANDARD MODEL
}
\author{Guido Altarelli \\
{\em Dipartimento di Matematica e Fisica, Universit\`a di Roma Tre;}\\
{\em INFN, Sezione di Roma Tre, }\\
{\em Via della Vasca Navale 84, I-00146 Rome, Italy}\\
{\em  and}  \\
{\em CERN, Department of Physics, Theory Unit,}\\
{\em CH-1211 Geneva 23, Switzerland}
}

\maketitle
\baselineskip=11.6pt
\begin{abstract}
The LHC runs at 7 and 8 TeV have led to the discovery of the Higgs boson at 125 GeV which will remain as one of the major physics discoveries of our time. Another very important result was the surprising absence of any signals of new physics that, if confirmed in the continuation of the LHC experiments, is going to drastically change our vision of the field. Indeed the theoretical criterium of naturalness required the presence of new physics at the TeV scale. At present the indication is that Nature does not too much care about our notion of naturalness. Still the argument for naturalness is a solid one and one is facing a puzzling situation. We review the different ideas and proposals that are being considered in the theory community to cope with the naturalness problem. 
\vskip .1cm
$~~~~~~~~~~~~~~~~~~~~~~~~~~~~~~~~~{\rm RM3-TH/14-12;~~~~~~CERN-PH-TH/2014-127}$
\end{abstract}
\baselineskip=14pt

\section{\bf Introduction}

With the discovery at the LHC \cite{ATLH,CMSH} of a particle that, in all its properties, appears just as the Higgs boson of the Standard Model (SM), the main missing block for the experimental validation of the theory is now in place. The Higgs discovery is the last milestone in the long history (some 130 years) of the development of  a field theory of fundamental interactions (apart from quantum gravity),
starting with the Maxwell equations of classical electrodynamics, going through the great revolutions of Relativity
and Quantum Mechanics, then the formulation of Quantum Electro Dynamics (QED) and the gradual build up of the gauge part of the Standard Model
and finally completed with the tentative description of the  Electro-Weak (EW) symmetry breaking sector of the SM in terms of a simple formulation of the Englert- Brout- Higgs mechanism \cite{ebh}. 

An additional LHC result of great importance is that a large new territory has been explored and no new physics was found. If one considers that there has been a big step in going from the Tevatron at 2 TeV up to the LHC at 8 TeV (a factor of 4) and that only another factor of  1.75 remains to go up to 14 TeV, the negative result of all searches for new physics is particularly depressing but certainly brings a very important input to our field with a big change
in perspective. In particular, while New Physics (NP) can still appear at any moment, clearly it is now less unconceivable that no new physics will show up at the LHC.  

As is well known, in addition to the negative searches for new particles,  the constraints on new physics from flavour phenomenology are extremely demanding:
when adding higher dimension effective operators to the SM, the flavour constraints generically lead  to powers of very large suppression scales $\Lambda$ in the denominators of the corresponding coefficients. In fact in the SM there are powerful protections against flavour changing neutral currents and CP violation effects, in particular through the smallness of quark mixing angles. Powerful constraints also arise from the leptonic sector. In particular, we refer to the recent improved MEG result \cite{meg} on the $\mu \rightarrow e \gamma$ branching ratio, $Br(\mu \rightarrow e \gamma) \leq 5.7\times10^{-13}$ at $90\%$ C.L. and to other similar processes like $\tau \rightarrow (e~\rm{or}~ \mu)  \gamma$ and to the bound on the electron dipole moment $|d_e| \lappeq 8.7~10^{-29}$ e cm by the ACME Collaboration \cite{ACME}.
In this respect the SM is very special and, as a consequence, if there is new physics, it must
be highly non generic in order to satisfy the present constraints. 

There is no evidence of new physics from accelerator experiments (except, perhaps, for the 3-3.5 $\sigma$ discrepancy of the muon (g-2) \cite{amuexp,amu}). Most of the experimental evidence for NP comes from the sky like for Dark Energy, 
Dark Matter, baryogenesis and also neutrino oscillations (that were first observed in solar and atmospheric neutrinos). 
One expected new physics at the EW scale  based on a "natural" solution of the hierarchy problem \cite{hiera}. The absence so far of new physics signals casts doubts on the relevance of our concept of naturalness. In the following we will elaborate on this naturalness crisis. 
  
\section{\bf The impact of the Higgs discovery}

A particle that, within the present accuracy, perfectly fits with the profile of the  minimal SM Higgs has been observed at the LHC. Thus, what was considered just as a toy
model, a temporary addendum to the gauge part of the SM, presumably to be replaced by a more complex reality and likely to be accompanied by new physics,
has now been experimentally established as the actual realization of the EW symmetry breaking (at least to a very good approximation).
It appears to be the only known example in physics of a fundamental,
weakly coupled, scalar particle with vacuum expectation value (VEV). We know many composite types of Higgs-like particles, like the Cooper pairs of superconductivity or the quark condensates that break the chiral symmetry of massless QCD, but the LHC Higgs is the only possibly elementary one.
This is a death blow not only to Higgsless models, to straightforward technicolor
models and other unsophisticated strongly interacting Higgs sector models but actually a threat to all models without
fast enough decoupling (in that if new physics comes in a model with decoupling the
absence of new particles at the LHC helps in explaining why large corrections
to the H couplings are not observed).
The mass of the Higgs is in good agreement with the predictions from the EW precision tests analyzed in the SM \cite{smfit}. The possibility of a "conspiracy" (the Higgs is heavy but it falsely appears as light because of confusing new physics effects) has been discarded: the EW precision tests of the SM tell the truth and in fact, consistently, no "conspirators", namely no new particles, have been seen around. 

\section{\bf Our concept of naturalness is challenged} 

The simplicity of the Higgs is surprising but even more so is the absence of accompanying new physics: this brings the issue
of the relevance of our concept of naturalness at the forefront.
As is well known, in the SM the Higgs provides a solution to the occurrence of
unitarity violations that, in the absence of a suitable remedy, occur in some amplitudes involving longitudinal gauge bosons as in $V_LV_L$ scattering, with $V=W, Z$ \cite{unit}.
To avoid these violations one needed either one or more
Higgs particles or some new states (e.g. new vector bosons).
Something had to happen at the few TeV scale!

While this prediction is based on a theorem, once there is a Higgs particle, the threat of unitarity violations is tamed and the
necessity of new physics on the basis of naturalness has not the same status, in the sense that it is not a theorem. The naturalness principle has been and still remains the main argument for new physics at
the weak scale. But at present our confidence on naturalness as a guiding
principle is being more and more challenged. Manifestly, after the LHC 7-8 results, a substantial amount of fine tuning is imposed 
on us by the data. So the questions are: does Nature really care about our concept of 
naturalness? Apparently not much! Should one give up naturalness? Which form of naturalness is natural?

The naturalness argument for new physics at the EW scale
is often expressed in terms of the quadratic cut-off dependence
in the scalar sector, before renormalization. If we see the cut-off  as the scale where new physics 
occurs that solves the fine tuning problem, 
then this new physics must be nearby because the observed scalar mass $m$ and the cut-off should a priori be of the same order  (modulo coupling factors). Actually the argument can be formulated in terms of renormalized
quantities, with no reference to a cut-off, but rather in terms of a
quadratic  sensitivity to thresholds at high energy. In the renormalized theory
the running Higgs mass $m$
slowly evolves logarithmically according to the relevant beta functions \cite{strubeta}. But in the presence of a threshold 
at $M$ for a heavy particle with coupling $\lambda_H$
to the Higgs, the quadratic 
sensitivity produces a jump in the
running mass of order $\Delta m^2 \sim (\lambda_H M)^2/16\pi^2$ (see, for example, \cite{barThr}).  In the presence of a threshold at $M$ one needs a fine tuning of order $m^2/M^2$ in order to reproduce the observed value of the running mass $m$ at low energy.

The argument for naturalness, although very solid in principle, certainly has failed so far as a guiding principle. As a consequence:
we can no more be sure that within 3 or 10 or 100 TeV..... 
the solution of the hierarchy problem must be found, 
which, of course, has negative implications for the design of future Colliders. 
Moreover, it is true that the SM theory is renormalizable 
and completely finite and predictive.
If you forget the required miraculous fine tuning you are 
not punished, you find no catastrophe!
The possibility that the SM holds well beyond the EW scale
must now be seriously considered.
The absence of new physics appears as a paradox to us. Still the picture repeatedly suggested by the data 
in the last ~20 years is simple and clear: take the SM, extended to include Majorana neutrinos and
some form of Dark Matter, as valid up to some very high energy. 

There is actually no strict argument that prevents to extend the validity of the SM at large energies. It turns out that the observed value of  the Higgs mass $m$ is a bit too low for the SM to be valid up to the Planck mass with an absolutely stable vacuum. The pure SM evolution of couplings, given the measured values of the top and Higgs masses and of the strong coupling $\alpha_s$, appears to
lead to a metastable Universe with a lifetime longer than the age of the Universe, so that the SM can well be valid up to the Planck mass (if one is ready to accept the immense fine tuning that this option implies). Also, it is puzzling to find that the evolution of the Higgs quartic coupling ends up into a narrow metastability wedge at very large energies. This result is obtained from a recent state-of-the-art evaluation of the relevant boundaries \cite{strubeta}. This criticality looks intriguing and perhaps it should tell us something. Note however that these results are obtained in the assumption of no new physics while possibly the solution of the Dark Matter problem or the presence of whatever new intermediate threshold could change the results. Actually also a peculiar behaviour of the Higgs potential near the Planck mass could alter the evaluation of the Universe lifetime \cite{branc}. 
Thus one cannot guarantee that the simplest picture is actually realized in detail but it is important that this possibility exists.

Thus, ignoring the implied huge fine tuning minimal extensions of the SM 
are being considered. Neutrino masses can be accommodated by introducing Right-Handed (RH) neutrinos and the See-Saw mechanism.
Baryogenesis, which represents a problem in the minimal SM, can be elegantly obtained through leptogenesis.
The solution of the crucial Dark Matter (DM) problem  could be in terms of some simple Weakly Interacting Massive Particle (WIMP), or by Axions, or by some keV sterile $\nu$Õs..... 
Coupling Unification without Supersymmetry (SUSY) could be restored by some large scale threshold, 
e.g. non-SUSY SO(10) with an intermediate scale (see, for example, Ref. \cite{melon}), and so on.
We now briefly discuss some of these possibilities.

\section{\bf Neutrino masses}

It is often stated that neutrino masses are the first observed form of NP. This is true but, in this case, a simple, elegant and conceptually far reaching extension of the SM directly leeds to an attractive framework for $\nu$ mass and mixing (for reviews, see Refs. \cite{revs}). It is sufficient to introduce 3 RH gauge singlets $\nu_R$,
each completing a 16 of $SO(10)$ for one generation,
and not artificially impose that the lepton number $L$ is conserved. We consider that the existence of RH neutrinos $\nu_R$ is quite plausible also because most GUT groups larger than SU(5) require
them. In particular the fact that $\nu_R$ completes the representation 16 of SO(10): 16=$\bar 5$+10+1, so that all
fermions of each family are contained in a single representation of the unifying group, is too impressive not to be
significant. At least as a classification group SO(10) must be of some relevance in a more fundamental layer of the theory! In the SM, in the absence of $\nu_R$, $B$ and $L$ are ÒaccidentalÓ 
symmetries, i.e. no renormalizable gauge invariant 
$B$ and/or $L$ non-conserving vertices can be built from 
the fields of the theory. But we know that, even in the absence of  $\nu_R$, non perturbative terms (instantons) 
break $B$ and $L$ (not $B-L$) and so do also non renormalizable operators like the Weinberg
dim-5 operator $O_5 = (Hl)^T_i\lambda_{ij}(Hl)_j/\Lambda$. With $\nu_R$ the Majorana mass term $M\nu^T_R \nu_R$ is allowed by $SU(2)\otimes U(1)$
($\nu_R$ is a gauge singlet!) and breaks $L$ (and $B-L$). A very natural and appealing description of neutrino masses can be formulated in terms of the see-saw mechanism \cite{seesaw}: the light neutrino masses are quadratic in the Dirac
masses and inversely proportional to the large Majorana mass.   Note that for
$m_{\nu}\approx \sqrt{\Delta m^2_{atm}}\approx 0.05$ eV and 
$m_{\nu}\approx m_D^2/M$ with $m_D\approx v
\approx 200~GeV$ we find $M\approx 10^{15}~GeV$ which indeed is an impressive indication for
$M_{GUT}$.

We have seen that in the presence of a threshold at $M$ one needs a fine tuning of order $m^2/M^2$ in order to reproduce the observed value of the running Higgs mass at low energy. Note that heavy RH neutrinos, which are coupled to the Higgs through the Dirac Yukawa coupling, would contribute in the loop and it turns out that, in the absence of SUSY,  become unnatural at $M \gappeq 10^7 - 10^8$ GeV \cite{nuunnat}. Also, in the pure Standard Model heavy $\nu_R$ tend to further destabilize the vacuum and make it unstable for $M \gappeq 10^{14}$ GeV \cite{nudest}.
 
The detection of neutrino-less double beta decay \cite{zuber} would provide direct evidence of $L$ non conservation and of the Majorana nature of neutrinos. It would also offer a way to possibly disentangle the 3 cases of degenerate, normal or inverse hierarchy neutrino spectrum. At present the best limits from the searches with Ge lead to $\vert m_{ee}\vert~\sim ~(0.25-0.98)$ eV (GERDA +HM +IGEX) and with Xe to $\vert m_{ee}\vert~\sim ~(0.12-0.25)$ eV (EXO +Kamland Zen), where ambiguities on the nuclear matrix elements lead to the ranges shown.
In the next few years, experiments (CUORE, GERDA II, SNO+....) will reach a larger sensitivity on $0\nu \beta \beta$ by about an order of magnitude. Assuming the standard mechanism through mediation of a light massive Majorana neutrino, if these experiments will observe a signal this would indicate that the inverse hierarchy is realized, if not, then the normal hierarchy case still would remain a possibility. 

\subsection{Baryogenesis via leptogenesis from heavy $\nu_R$ decay}

In the Universe we observe an apparent excess of baryons over antibaryons. It is appealing that one can explain the
observed baryon asymmetry by dynamical evolution (baryogenesis) starting from an initial state of the Universe with zero
baryon number.  For baryogenesis one needs the three famous Sakharov conditions: B violation, CP violation and no thermal
equilibrium. In the history of the Universe these necessary requirements have probably occurred together several times at different epochs. Note
however that the asymmetry generated during one such epoch could be erased in following epochs if not protected by some dynamical
reason. In principle these conditions could be fulfilled in the SM at the electroweak phase transition. In fact, when kT is of the order of a few TeV, B conservation is violated by
instantons (but B-L is conserved), CP symmetry is violated by the Cabibbo-Kobayashi-Maskawa phase and
sufficiently marked out-of- equilibrium conditions could be realized during the electroweak phase transition. So the
conditions for baryogenesis  at the weak scale in the SM superficially appear to be fulfilled. However, a more quantitative
analysis
\cite{kaj,tro} shows that baryogenesis is not possible in the SM because there is not enough CP violation and the phase
transition is not sufficiently strong first order, because the Higgs mass is too heavy. In SUSY extensions of the SM, in particular in the minimal SUSY model (MSSM), there are additional sources of CP violation but also this possibility has by now become at best marginal after the results from LEP2 and the LHC.

If baryogenesis at the weak scale is excluded by the data still it can occur at or just below the GUT scale, after inflation.
But only that part with
$|{\rm B}-{\rm L}|>0$ would survive and not be erased at the weak scale by instanton effects. Thus baryogenesis at
$kT\sim 10^{10}-10^{15}~{\rm GeV}$ needs B-L violation and this is also needed to allow $m_\nu$ if neutrinos are Majorana particles.
The two effects could be related if baryogenesis arises from leptogenesis then converted into baryogenesis by instantons
\cite{fuku,buch}. The decays of heavy Majorana neutrinos (the heavy eigenstates of the see-saw mechanism) happen with non conservation of lepton number L, hence also of B-L and can well involve a sufficient amount of CP violation. Recent results on neutrino masses are compatible with this elegant possibility. Thus the case of  baryogenesis through leptogenesis has been boosted by the recent results on neutrinos.

\section{\bf Dark Matter}

At present Dark Matter (DM) is the crucial problem. There is by now a robust evidence for DM in the Universe from a variety of astrophysical and cosmological sources. While for neutrino masses and baryogenesis, as we have seen,
there are definite ideas on how these problems could be solved,
DM remains largely mysterious and is a very
compelling argument for New Physics and the most pressing 
challenge for particle physics.

The 3 active $\nu$Õs cannot make the whole of DM. Nearby sterile $\nu$Õs  with $m_{\nu} \sim $ eV are also inadequate.  Bounds from
 dwarf galaxies require that $m_{\nu}\gappeq$ few hundreds eV (Tremaine-Gunn), from 
 galaxies $m_{\nu}\gappeq$ few tens eV.
 Hot DM (like neutrinos) is also excluded by structure formation.

WIMPS with masses in the range $10^{-1} - 10^3$ GeV and electroweak cross-sections remain optimal candidates. For WIMPÕs in thermal equilibrium after inflation the relic density can reproduce the observed value for typical EW cross-sections. This ÒcoincidenceÓ is taken as a good indication in favour of a WIMP explanation of DM. In SUSY models with R-parity conservation  the neutralino is a very attractive candidate for a WIMP (in SUSY also other candidates are possible like the gravitino). At the LHC there is a great potential for discovery of most kinds of WIMPÕs. So far no WIMPÕs have been observed at the LHC. But the LHC limits on neutralinos are not stringent: in large regions of parameter space $m_{\chi} \lappeq$ 350 GeV is allowed. A strict bound is very low: $m_{\chi} \gappeq$ 25 GeV (with light s-taus and higgsinos) \cite{calibbi}. 

The WIMP non-accelerator search continues and is very powerful (LUX, XENON, CDMS.....). The limits are generally given in a plane of mass versus cross-section (either spin dependent or spin independent) for processes like (for example, for fermionic DM $\chi$) $\chi +N \rightarrow \chi +N$ or $\chi +\chi \rightarrow N+\bar{N}$ with $N$ a nucleon. These processes could go via $Z$ exchange (among SM particles) and the present limits exclude a large range of $Z\chi \chi$ couplings for typical WIMP masses \cite{desim}. The axial couplings are the least constrained. Another possible SM mediator is the Higgs boson. Here the limits are less stringent, in particular for a pseudoscalar coupling  \cite{desim}. At present we can state that there is still plenty of room for WIMPÕs especially at low masses ($\sim$ 10 -100 GeV), or at large masses $\sim$ 1 - 10 TeV).

A rather minimal explanation for DM could be provided by axions, introduced originally to solve the strong CP problem \cite{peccei}. For a viable axion model some new particles that carry the $U(1)_{PQ}$ charge must exist at a scale $f$, for example some fermions $\Psi$ and a scalar $A$ \cite{pequi,kim,kimcp}. A part from the chiral anomaly, the $U(1)_{PQ}$ symmetry is broken by the $A$ VEV of order $f$, which also gives a mass to $\Psi$ of the same order of magnitude modulo some Yukawa-like coupling. The phase of $A$ is the axion field $a$ which is the Goldstone boson associated with the breaking of $U(1)_{PQ}$. It would be massless and only derivatively coupled if not for the chiral anomaly that gives a mass to the axion, inversely proportional to $f$, hence very small. The typical window for an axion that could explain the observed relic density is $f \sim 10^{10}- 10^{11}$ GeV and $m_a \sim  10^{-4}- 10^{-5}$ eV. The chiral anomaly also induces the decay $a \rightarrow \gamma \gamma$, through which the axion can be observed.  Clearly experimental axion searches are very important. So far the experiments were not sensitive enough to probe the relevant ranges of $f$ and $m_a$.  Now the Axion Dark Matter Experiment  (ADMX) plans to reach the required sensitivity in the next few years. 

\section{\bf Theory confronts the naturalness riddle}

To cope with the naturalness riddle different lines of thought have emerged. Here is a partial list:

1) Insist on minimizing the fine tuning (FT) within the present experimental constraints. In practice this amounts to imagine
suitable forms of new physics at an energy scale as close as possible (with new particles that could hopefully be observable at the LHC14).

2) Accept FT  only up to a large intermediate scale (i.e. still far below $M_{GUT}$):
e.g. split SUSY.

3) Make the extreme choice of a total acceptance of FT : 
the most typical approach being the anthropic philosophy.

4) Argue that possibly there is no FT :
make the conjecture that there is no new threshold  up to $M_{Pl}$ and invoke some miracle within the theory of quantum gravity to solve the naturalness of the EW versus the Planck scale.

We now briefly comment on these different options.

The first possibility is the most conservative and consists of continuing all efforts to minimize the FT. The goal is to implement some form of ''Stealth Naturalness'': build models where naturalness is 
restored not too far from the weak scale but the related
NP is arranged to be not visible so far. Those are clearly the best scenarios for the next LHC runs! The risk is to end up with baroque models where one is fine-tuning the fine-tuning-suppression 
mechanism. The two main directions along these lines are SUSY and Composite models. 
On the SUSY side, which, except for its most minimal versions, still remains the best NP framework, the simplest new ingredients for an orderly retreat \cite{natsusy} are compressed spectra,  heavy first two generations and the next-to-minimal NMSSM \cite{nmssm} (with an additional Higgs singlet).  These attempts represent the last trench of natural SUSY. 
In composite Higgs models \cite{compoold,compo,extop} naturalness is improved by the pseudo-Goldstone nature of the Higgs. However, minimal fine tuning demands the scale of compositeness $f$ to be as close as possible, or the $\xi = v^2/f^2$ parameter to be as large as possible ($v$ being the HIggs VEV). But this is limited by EW precision tests that demand $\xi <0.05-0.2$. Also the measured Higgs couplings interpreted within composite models lead to upper bounds on $\xi$. While in SUSY models the quadratic sensitivity of the top loop correction to the Higgs mass is quenched by a scalar particle, the s-top, in composite Higgs models the cancelation occurs with a fermion, either with the same charge as the top quark or even with a different charge. For example the current limit from a search of a $T_{5/3}$ fermion of charge 5/3 is $M_{T5/3} \geq 750$ GeV  \cite{CMST5/3} (an exotic charge quark cannot mix with ordinary quarks: such mixing would tend to push its mass up).

Given that our concept of naturalness has so far failed,   
there has been a revival of models that ignore the fine tuning problem while trying to accommodate the known facts.  For example, several fine tuned SUSY extensions of the SM have been studied  like Split SUSY \cite{split} or High Scale SUSY \cite{lssusy,giustru}. 
There have also been reappraisals of non SUSY Grand Unified Theories (GUT) where again one completely disregards fine tuning 
\cite{so10,ax,melon}. 
In Split SUSY only those s-partners are light that are needed for Dark Matter and coupling unification, i.e. light gluinos, charginos and neutralinos (also A-terms are small) while all scalars are heavy (a hierarchy explained in terms of a chiral symmetry or a discrete parity). As a result also flavour problems are very much eased down. The measured Higgs mass imposes an upper limit to the large scale of heavy s-partners \cite{giustru} which, for Split SUSY, is at $10^4 - 10^7$ GeV, depending on $\tan{\beta}$, while in High-Scale SUSY, where all supersymmetric partners have roughly equal masses of order $M_{SUSY}$, the latter must fall in the range $10^3 - 10^{10}$ GeV, again depending on $\tan{\beta}$. It is interesting that in both cases the value of $M_{SUSY}$ must be much smaller than $M_{GUT}$. In both Split SUSY and High-Scale SUSY the relation with the Higgs mass occurs through the quartic Higgs coupling, which in a SUSY theory is related to the gauge couplings. In turn the quartic coupling is  connected to the Higgs mass via the minimum condition for the Higgs potential.  In Split SUSY it is not granted but still possible that the light gluinos, charginos and neutralinos can be observed at the LHC.

An extreme point of view (but not excluded) is the anthropic evasion of the problem, motivated by the fact that the observed value of the cosmological constant $\Lambda_{cosmo}$ also poses a tremendous, unsolved naturalness problem \cite{tu}. Yet the value of $\Lambda_{cosmo}$ is close to the Weinberg upper bound for galaxy formation \cite{We}. Possibly our Universe is just one of infinitely many bubbles (Multiverse) continuously created from the vacuum by quantum fluctuations (based on the idea of chaotic inflation). Different physics takes place in different Universes according to the multitude of string theory solutions ($\sim 10^{500}$ \cite{doug,shellek}). Perhaps we live in a very unlikely Universe but the only one that allows our existence \cite{anto},\cite{giu}. Given the stubborn refusal of the SM to show some failure and the
terrible unexplained naturalness problem of the 
cosmological constant, many people have turned to the
anthropic philosophy also for the SM. Actually applying the anthropic principle to the SM hierarchy problem is not so convincing. After all, we can find plenty of models that reduce the fine
tuning from $10^{14}$ down to $10^2$.  And the added ingredients apparently
would not make our existence less possible.
So why make our Universe so terribly unlikely? Indeed one can argue that the case of the cosmological constant 
is a lot different: the context is not as fully specified  as the for the SM. Also so far there is no natural theory of the cosmological constant. On the other hand there is some similarity:
$\Lambda_{cosmo}$ corresponds to a vacuum energy density in all points of space
just like the Higgs VEV $v$ (which actually makes a contribution to $\Lambda_{cosmo}$ that must be mysteriously canceled). With larger $\Lambda_{cosmo}$ there is no galaxy formation, with larger $v$ no nuclear physics. The anthropic way is now being kept in mind as a possibility.

We have seen that the hierarchy problem is manifested by the quadratic sensitivity of  the scalar sector mass scale $m$ to the physics at large energy scales. In the presence of a threshold at $M$ one needs a fine tuning of order $m^2/M^2$ in order to reproduce the observed value of the running Higgs mass $m$ at low energy. A possible point of view is that there are no new thresholds up to $M_{Planck}$ (at the price of giving up GUTs, among other things) but, miraculously, there is a hidden mechanism in quantum gravity that solves the fine tuning problem related to the Planck mass \cite{shapo,gian}. For this one would need to solve all phenomenological problems, like DM, baryogenesis and so on, with physics below the EW scale.  This point of view is extreme but allegedly not yet ruled out. In this context the sensational announcement by the BICEP2 Collaboration \cite{bicep} of the observation of a rather large value of the ratio $r$ of tensor to scalar polarization modes in the Cosmic Microwave Background, $r \sim 0.2 \pm ^{0.07}_{0.05}$. This result would imply an energy scale of inflation given by $V_{infl}^{1/4} \sim 2.2~10^{16}~(r/0.2)^{1/4}$ (note the fourth root that makes this energy scale rather insensitive to the precise value of $r$).  The coincidence of this energy scale with $M_{GUT}$ is really amazing. For the implications of the BICEP2 results on axion masses and couplings, see Refs. \cite{vis,dival}. It must be stressed that the BICEP2 claim needs to be confirmed by new data, also in view of widespread doubts on the procedure of subtraction of the dust foreground \cite{flaug}. 

Possible ways to realize the no threshold program are discussed in Ref. \cite{shapo}: one has to introduce three RH neutrinos, $N_1$, $N_2$ and $N_3$ which are now light: for $N_1$ we need $m_1$ ~few keV, while $m_{2,3}$ ~ few GeV but with a few eV splitting. With this rather ad hoc spectrum $N_1$ can explain DM and $N_{2,3}$ baryogenesis. The active neutrino masses are obtained from the see-saw mechanism, but with very small Dirac Yukawa couplings. Then the data on neutrino oscillations can be reproduced. The RH $N_i$ can give rise to observable consequences (and in fact only a limited domain of the parameter space is still allowed). In fact $N_1$ could decay as $N_1 \rightarrow \nu + \gamma$ producing a line in X-ray spectra at $E_\gamma \sim m_1/2$. It is interesting that a candidate line with $E_\gamma \sim 3.5$ keV has been identified in the data of the XMM-Newton X-ray observatory on the spectra from galaxies or galaxy clusters \cite{line}.  As for $N_{2,3}$ they could be looked for in charm meson decays if sufficiently light. A Letter of Intent for a dedicated experiment at the CERN SpS has been presented to search for these particles \cite{loi}. 

In this class of theories one can also mention a more restrictive dynamical possibility: scale invariant theories 
possibly including gravity (see \cite{agrav} and Refs. therein) where
only a-dimensional couplings exist and there is a 
spontaneous breaking of scale invariance. The problem, not surprisingly, is to explain the two very different scales of symmetry breaking at the EW and the Planck scale.

\section{\bf Summary and conclusion}

Among the main results at the LHC7-8 have been the discovery of a Higgs boson that, within the limits of the present, not too precise, accuracy, very much looks as minimal, elementary and standard and the absence of any direct or indirect signal of accompanying NP, which was expected on the basis of naturalness. Apparently our naive notion of naturalness has failed as a heuristic principle. We can say that we expected complexity and instead we have found a maximum of simplicity. Of course there are strong empirical evidences for NP beyond the SM that mostly arise not from accelerators but, one could say, from the sky, like Dark Energy, DM,  baryogenesis and neutrino masses.  But the picture repeatedly suggested by the data 
in the last ~20 years is simple and clear: take the SM, extended to include Majorana neutrinos, which can explain the smallness of active neutrino masses by the see-saw mechanism and baryogenesis through leptogenesis, plus
some form of DM, as valid up to some very high energy. Indeed at present in particle physics the most crucial experimental problem is the nature of DM. In this case a vast variety of possible solutions exist from WIMPS to axions or to keV sterile neutrinos or.... Clearly which of the many possible solutions or which combination of them will eventually be established will impose a well definite path for going beyond the SM. We have discussed a number of approaches to confront the naturalness riddle, including insisting on minimizing the fine tuning (FT) within the present experimental constraints or accepting FT  only up to a large intermediate scale but still far below $M_{GUT}$), like for split SUSY or making the extreme choice of a total acceptance of FT: 
as in the anthropic point of view or arguing that possibly there is no FT with no new threshold  up to $M_{Pl}$ and invoking some miracle within the theory of quantum gravity (at the price of giving up Grand Unification and heavy RH neutrinos below the Planck scale). Clearly we are experiencing a very puzzling situation but, to some extent, this is good because big steps forward in fundamental physics have often originated from paradoxes. We highly hope that the continuation of the LHC experiments will bring new light on these problems.

\section{\bf Acknowledgements}
I am very grateful to the Organizers (in particular to Profs. Gianpaolo Mannocchi and Roberto Fusco-Femiano) for their invitation and hospitality. This work has been partly supported by the Italian Ministero dell'Uni\-ver\-si\-t\`a e della Ricerca Scientifica, under the COFIN program (PRIN 2008), by the European Commission, under the networks ``LHCPHENONET'' and ``Invisibles''.


%
\end{document}